\documentclass[a4paper,
               keeplastbox,   % flushend option: not to un-indent last line in References
               ]{jacow}
\usepackage{pdfpages,multirow,ragged2e} %
\makeatletter%
	\ifboolexpr{bool{xetex}}
	 {\renewcommand{\Gin@extensions}{.pdf,%
	                    .png,.jpg,.bmp,.pict,.tif,.psd,.mac,.sga,.tga,.gif,%
	                    .eps,.ps,%
	                    }}{}
\makeatother

\ifboolexpr{bool{xetex} or bool{luatex}} % test for XeTeX/LuaTeX
 {}                                      % input encoding is utf8 by default
 {\usepackage[utf8]{inputenc}}           % switch to utf8

\usepackage[USenglish]{babel}

\ifboolexpr{bool{jacowbiblatex}}%
 {%
  \addbibresource{jacow-test.bib}
  \addbibresource{biblatex-examples.bib}
 }{}
\begin{document}

\title{Installation, Commissioning and Performance of Phase Reference Line for LCLS-II\thanks{This work was supported by the LCLS-II Project and the U.S. Department of Energy, Contract DE-AC02-76SF00515 }}

\author{S.D. Murthy\thanks{sdmurthy@lbl.gov}, L. Doolittle,  LBNL, Berkeley, CA 94720, USA\\
 C. Xu, B. Hong, A. Benwell, J. Chen,  SLAC, Stanford, CA 94309, USA }

\maketitle

\begin{abstract}
Any cavity controller for a distributed system needs a Phase Reference Line (PRL) signal from which to define phases of a cavity field measurement. The LCLS-II PRL system at SLAC provides bidirectional (forward and reverse) phase references at 1300\thinspace MHz to each rack of the LLRF system. The PRL controller embedded with the Master Oscillator (MO) locks the average phase of the two directions to the MO itself. Phase-averaging tracking loop is applied in firmware which supports the feature of cancelling the phase drift caused by changes in PRL cable length. FPGA logic moves the phase of digital LO to get zero average phase of the two PRL signals. This same LO is used for processing cavity pickup signals, thus establishing a stable reference phase for critical cavity RF measurements. At low frequencies, open-loop PRL noise relative to the LO distribution includes a strong environment and $1/f$ components, but the closed-loop noise approaches the noise floor of the DSP. The implication is that the close-in phase noise of the cavities will be dominated by the chassis DAQ noise. The final out-of-loop phase noise relevant to machine operations is that of the cavity field relative to the beam.
\end{abstract}

\section{INTRODUCTION}
To achieve their physics-driven goals, all accelerator RF sub-systems need a stable phase reference. The primary job of the LLRF system is to hold the amplitude and phase of each cavity’s voltage constant, where cavity phase is defined relative to the phase reference signal. The LLRF hardware receives two frequencies, LO (1320\thinspace) and PRL (1300\thinspace MHz). Both of these signals are generated by the MO subsystem. The specifications for cavity field stability in LCLS-II are stringent by historical accelerator standards: 0.01$^\circ$ in phase and $0.01\%$ in amplitude \cite{req}. Careful analysis and design minimizes cavity phase noise and drift.

A phase averaging reference line scheme has been used in the field before at both SLAC \cite{IEEE} and Fermilab \cite{IEEE1}, to compensate for thermal drift in the cable used to distribute phase information. The PRL design adopted for LCLS-II starts from the same concept but uses digital techniques to perform the averaging.
When averaging is done using analog components, expert field setup is needed because of sensitivity to amplitude imbalance.
The LCLS-II system at SLAC provides bidirectional (forward and reverse) references to each rack of the LLRF system at 1300\thinspace MHz. By digitizing the two signal separately, more diagnostics are possible, and the equivalent adjustments performed by software.

For every offset frequency from DC up to the closed-loop cavity bandwidth, the LLRF system locks the cavity phase using combination of LO and PRL sources.
At low frequencies (below a few kHz), the PRL contribution dominates, so the information it provides (after averaging) needs to have minimum phase drifts \cite{lclsii}.

\section{DESIGN}
\subsection{Phase Reference System}
\begin{figure}[!htb]
   \centering
   \includegraphics*[width=0.9\columnwidth]{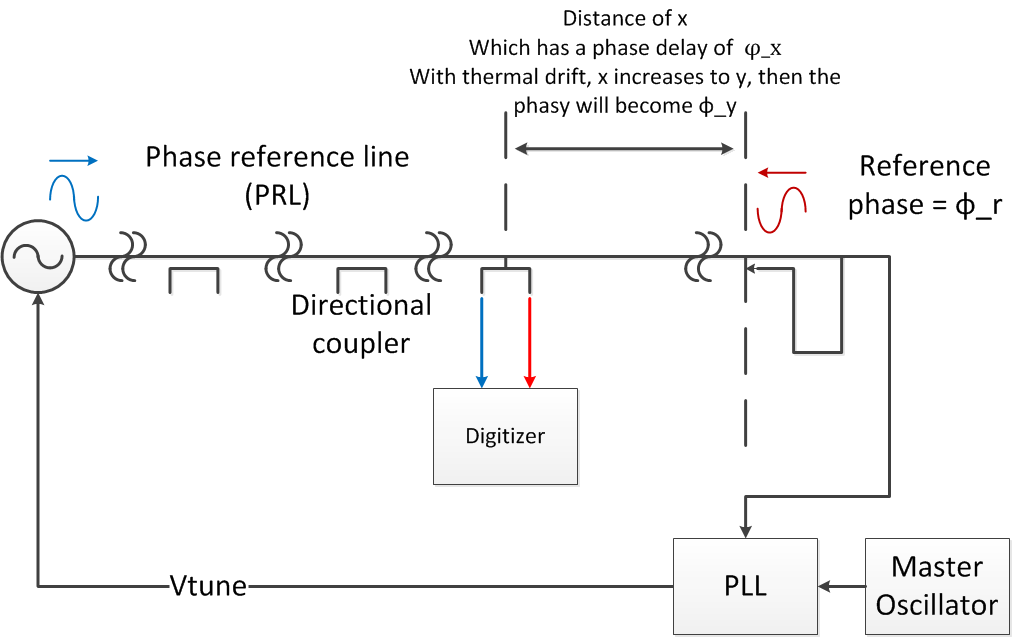}
   \caption{LCLS-II phase averaging technique.}
   \label{fig:block_diagram}
\end{figure}

A Voltage Controlled Oscillator (VCO) drives the remote end (far from the MO) of the physical PRL cable to generate a forward phase reference signal at 1300\thinspace MHz.
In the MO rack, that forward signal is actively reflected to generate the reverse wave on the physical PRL cable. A phase lock circuit pins the average of those two signals to the phase of the MO itself, creating the reflectometer technique \cite{IEEE2}. The block diagram of the phase averaging portion of the PRL is shown in Fig.~\ref{fig:block_diagram}. At any arbitrary point on the phase reference line, the directional coupler couples out the forward and reverse signals. The distance from the coupler to the MO determines the amount of phase difference. The forward and reverse signals are sent to the LLRF chassis to perform digital phase averaging.

\subsection{Installation}

The overall LCLS-II linac with 35 1.3\thinspace GHz cryomodules, spread out over 600 meters, is divided into segments, from L0 near the gun (a single cryomodule) to the long (20 cryomodules) L3 for the final on-crest energy boost. LH refers to the 3.9\thinspace GHz energy linearizer (two cryomodules). The PRL is divided up into three segments: L0+L1+LH, L2, and L3. The MO and most PRL support electronics are located between LH and L2. The phase-averaging feature described here is only used in the L0+L1+LH and L2 segments. The unidirectional L3 segment is driven based on an analog phase averager tap on the L2 PRL. Out-of-loop diagnostics can measure the consistency between the L0+L1+LH and L2 segments.

\begin{figure}[!htb]
   \centering
   \includegraphics*[width=1.0\columnwidth]{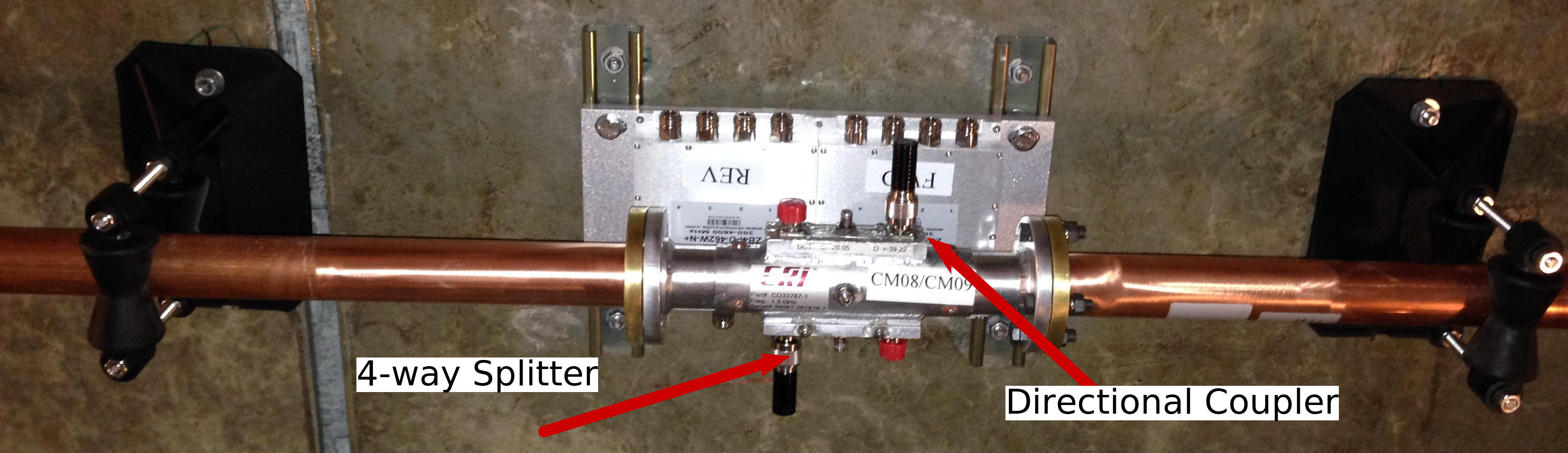}
   \caption{Physical PRL installation.}
   \label{fig:installation}
\end{figure}

As part of the original (2016) system design, the coaxial cable connection from the PRL coupler to the LLRF rack and the cable connection from the cryomodule to the LLRF racks are installed to be from the same cable roll with matching electrical lengths. This is so that they exhibit the same amount of phase drift due to temperature swings. Also, Time-Domain Reflectometer (TDR) is used to check any faults in these cables. Fig.~\ref{fig:installation} shows the physical installation of the PRL at cryomodules 8 and 9.

During this setup, PRL amplitude was attenuated to approximately -10\thinspace dBFS at each chassis to reduce the effect of any crosstalk at the digitizer.

\subsection{Cavity Phase Measurement Firmware}

A phase-averaging tracking loop is applied in firmware to eliminate the phase drift caused by the change in PRL cable length. Both the forward and reverse reference signals are first downconverted and digitized by the LLRF system, and further digitally downconverted to baseband.  These complex-number values are multiplied by the PRL gains and summed together to get an average phase, canceling drifts as previously described. This phase is integrated and passed to the digital downconverter shared between cavity and PRL signals, forming a tracking phase-locked loop.
The phase-averaging module can be easily enabled and disabled under software control. When it is disabled, the reference phase is driven to zero, and the system operates using the LO as a virtual phase reference. The above computation of phase-averaging adds four cycles of pipelining delays from the input to the averaged phase and uses a total of 2 multipliers (one for each pair of input) and 2 adders (adding each product and integrator).

The PRL gain registers are calculated and set by the software, based on the tracking bandwidth, nominal phase offset, actual phase offset, forward and reverse PRL signal strengths. The reason to have a nominal offset is that the setup process can be able to consistently choose one of the two possible values (180$^\circ$ apart) for forward/reverse offset that will bring those two signals into phase alignment. Either selection will work, but it's important to choose the same one across reboots. The actual phase offset is the final phase offset that will be subtracted and added to the forward and reverse reference signals respectively.  The forward and reverse PRL amplitudes along with the bandwidth are used to calculate the gain amplitudes. Finally, the PRL gain values are calculated based on the gain amplitude and the actual phase offset.

This firmware setup applies only to the L0, L1, LH, and L2 sections of the Linac. The L3 system is simpler, since there's only one PRL input to each rack. There is no need for any intricate offset considerations. To simplify deployment, the same firmware is used in both cases. In L3, we just lock to the single input (forward) with zero phase. The two PRL gain registers handling reverse signals get set to zero, and the two others depend only on the forward signal strength and desired tracking bandwidth.

\section{Commissioning and Performance}
The commissioning of the LCLS-II SRF system is currently ongoing at SLAC. Changes to firmware/software while testing on the real PRL cables is inevitable. One such change included adding guard bits to the phase-averaging firmware module to reduce rounding errors.
\begin{figure}[!htb]
   \centering
   \includegraphics*[width=1.1\columnwidth]{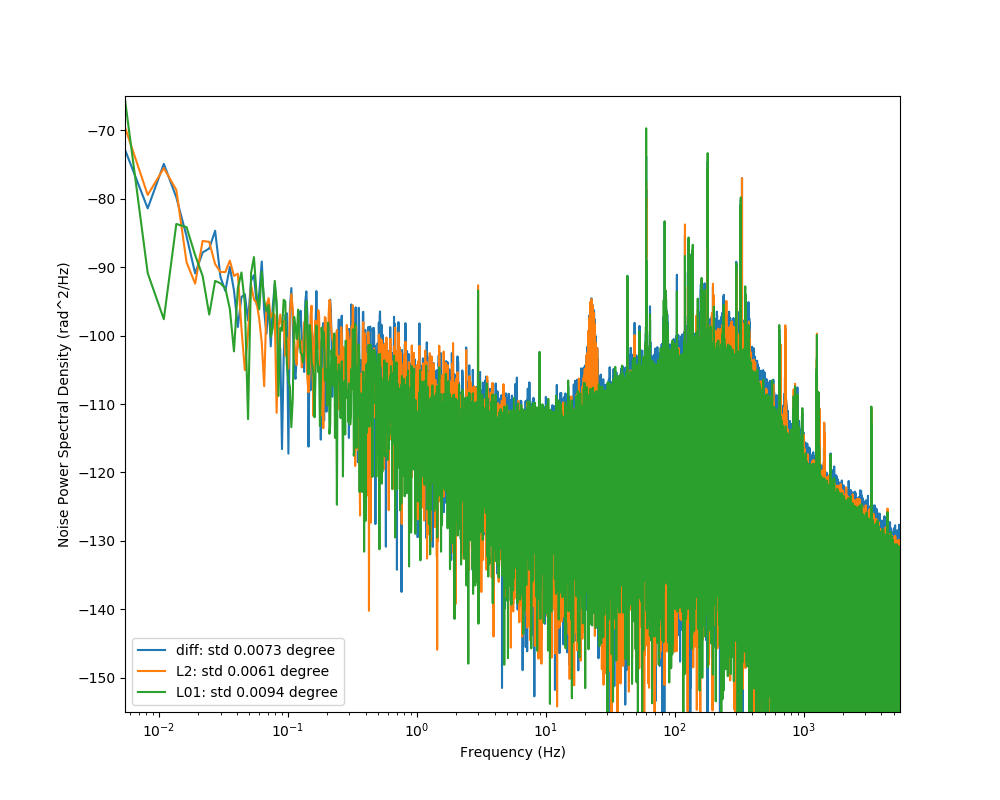}
   \caption{Phase noise power spectral density.}
   \label{fig:psd}
\end{figure}

\begin{figure}[!htb]
   \centering
   \includegraphics*[width=1.1\columnwidth]{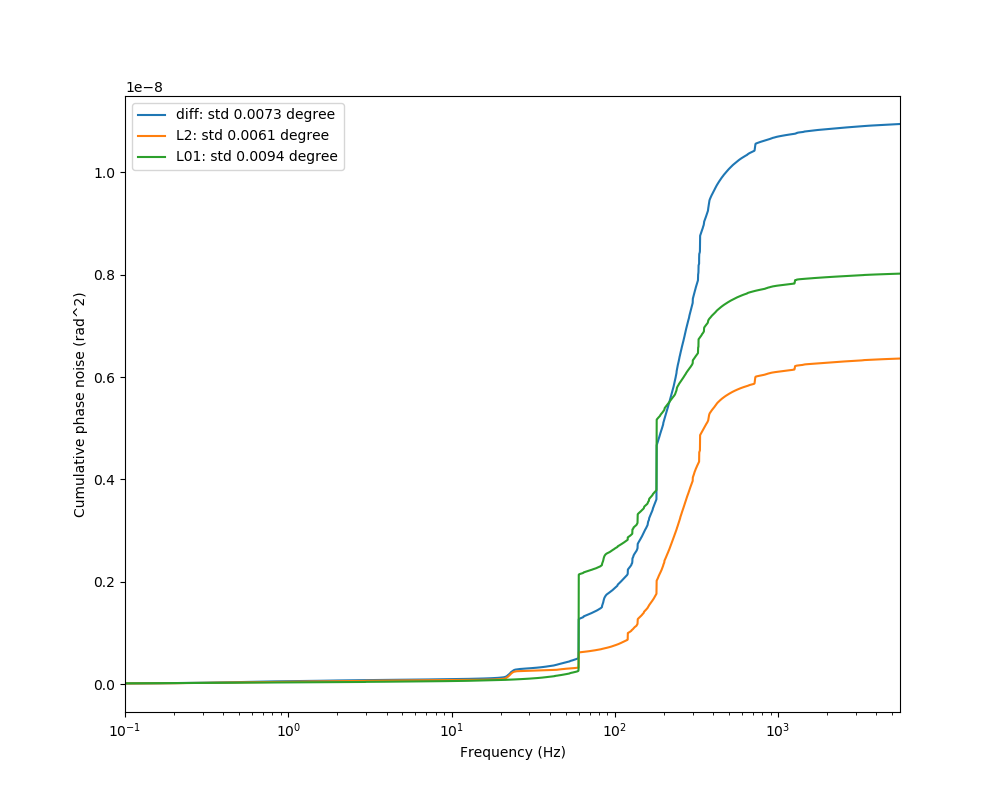}
   \caption{Cumulative phase noise plot.}
   \label{fig:cum}
\end{figure}

\begin{figure}[!htb]
   \centering
   \includegraphics*[width=1.0\columnwidth]{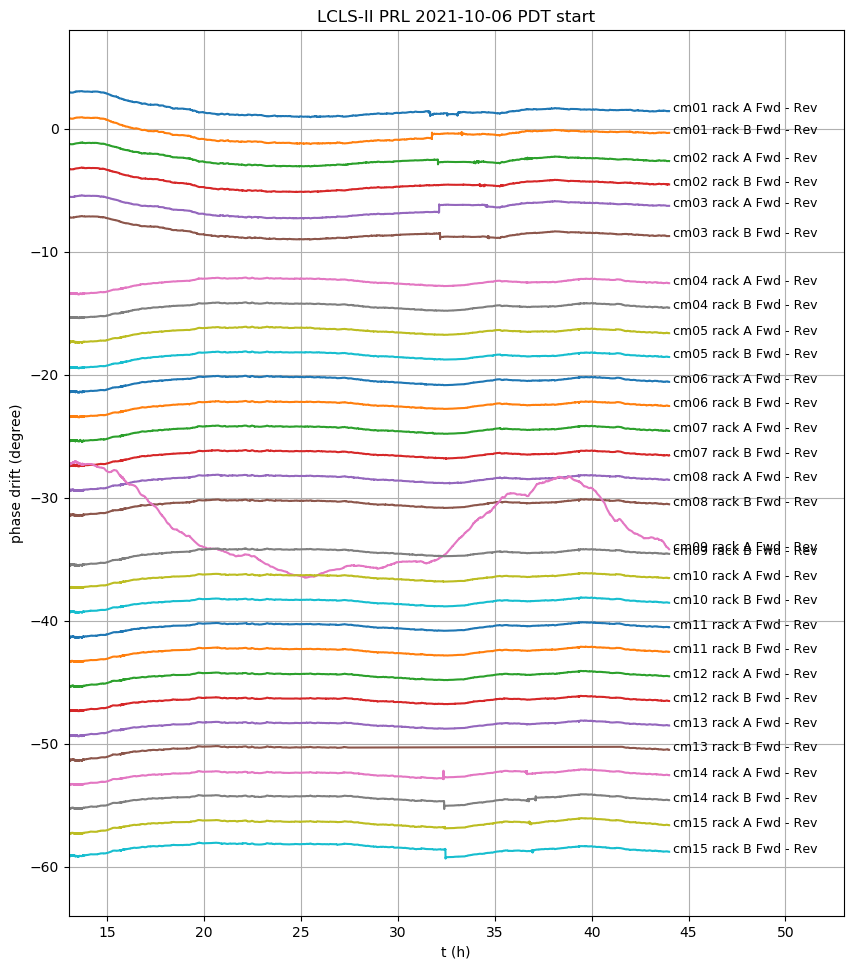}
   \caption{Phase drift of cables over the long term.}
   \label{fig:drift}
\end{figure}

A LLRF chassis was installed in the MO rack to monitor the upstream and downstream signals of each of the PRLs, so four RF signals in total (L0+L1 forward and reverse signals, L2 forward and reverse signals).
The jitter between the two reference lines is 0.0073$^\circ$ rms, with energy between 60\thinspace Hz and 1\thinspace kHz as shown in Fig.~\ref{fig:psd}.
The rms jitter on these individual reference lines L0+L1 and L2 are 0.0061$^\circ$ and 0.0094$^\circ$ respectively. In the cumulative phase noise plot shown in Fig.~\ref{fig:cum}, an artificial 0.1\thinspace Hz high-pass filter is applied before integrating, to avoid pathologies at DC. Due to the physical location of this chassis and the supporting couplers and cables, many error terms will be smaller here than in the "real" LLRF racks. This includes the noise from the PRL and noise of the measurement system itself.
Fig.~\ref{fig:drift} shows how the PRL cables to the LLRF chassis stretch over the long term due to temperature fluctuations. A defective coupler (since replaced) for cryomodule 9 rack A shows unusually large phase drifts.

\begin{figure}[!htb]
   \centering
   \includegraphics*[width=1.15\columnwidth]{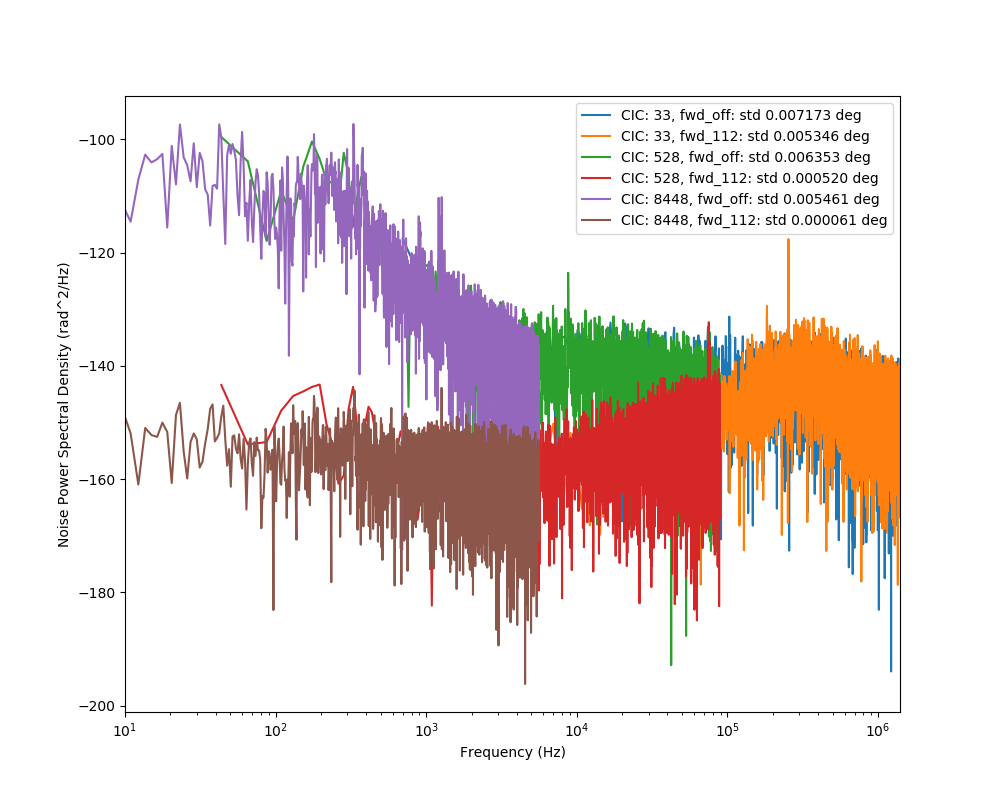}
   \caption{Noise spectrum of the phase-averaging tracking loop (open and closed-loop).}
   \label{fig:phs_noise}
\end{figure}

The maximum achievable tracking loop bandwidth is 300\thinspace kHz (limited by the firmware design), and the system has been tested at 10\thinspace kHz. Acquired noise spectra show the transition between high frequencies (PRL noise is unchanged) to low frequencies (noise is reduced by the PRL gain). At low frequencies, open-loop PRL noise relative to the LO distribution system includes a clear 1/f component. At low frequencies, closed-loop noise (in-loop error) approaches -150\thinspace dBrad$^2$/Hz representing the noise floor of the DSP as shown in Fig.~\ref{fig:phs_noise}.

\section{CONCLUSION}
The phase reference line was successfully installed at the LCLS-II to provide a stable reference phase for the LLRF system measurements. The (correctable) phase drift of the PRL cables over a long period has a span of about 2$^\circ$.
At low frequencies, closed-loop noise (in-loop error) approaches the noise floor of the DSP, and the measured jitter between the two reference lines easily meets the performance requirements.
The full out-of-loop performance measurement of the system will only be possible further along in the accelerator commissioning, when the phase noise of the beam relative to cavities can be measured.
%
% only for "biblatex"
%
\ifboolexpr{bool{jacowbiblatex}}%
	{\printbibliography}%
	{%
	% "biblatex" is not used, go the "manual" way

	%\begin{thebibliography}{99}   % Use for  10-99  references
	
} % end \ifboolexpr
%
% for use as JACoW template the inclusion of the ANNEX parts have been commented out
% to generate the complete documentation please remove the "%" of the next two commands
%
%%%\newpage

%%%\include{annexes-A4}

\end{document}